\documentclass[journal]{IEEEtran}
\usepackage{cite}
\usepackage{amsmath,amssymb,amsfonts}
\usepackage{algorithmic}
\usepackage{graphicx}
\usepackage{textcomp}
\def\BibTeX{{\rm B\kern-.05em{\sc i\kern-.025em b}\kern-.08em
    T\kern-.1667em\lower.7ex\hbox{E}\kern-.125emX}}
\usepackage[T1]{fontenc}
\usepackage{comment}
\usepackage{color}
\usepackage{bigints}
\usepackage{balance}
\usepackage{array}  
\usepackage{multirow}
\usepackage{graphicx}
\usepackage{amsmath}
\usepackage{pict2e}
\usepackage[cmintegrals]{newtxmath}
\usepackage{cite}

\ifCLASSINFOpdf
  
\else
  
\fi

\begin{document}

\title{A New RIS Architecture with a Single Power Amplifier: Energy Efficiency and Error Performance Analysis}

\author{Recep~A.~Tasci,~\IEEEmembership{Graduate~Student~Member,~IEEE,}
        Fatih~Kilinc,~\IEEEmembership{Graduate~Student~Member,~IEEE,}
        Ertugrul~Basar,~\IEEEmembership{Senior~ Member,~IEEE}
        and~George~C.~Alexandropoulos,~\IEEEmembership{Senior~ Member,~IEEE}
\thanks{Recep A. Tasci, Fatih Kilinc, Ertugrul Basar are with Communications Research and Innovation Laboratory (CoreLab), Department of Electrical and Electronics Engineering, Ko\c{c} University, 34450 Istanbul, Turkey (e-mail: \{ratasci20, fkilinc20, ebasar\}@ku.edu.tr}
\thanks{George C. Alexandropoulos is with Department of Informatics and Telecommunications, National and Kapodistrian University of Athens, 15784 Athens, Greece (email: alexandg@ieee.org)}
\thanks{Manuscript received December 20, 2021; revised March 15, 2022.}}

\maketitle

\begin{abstract}
Reconfigurable intelligent surface (RIS)-assisted communication have recently attracted the attention of the wireless communication community as a potential candidate for the next $6$-th generation (6G) of wireless networks. Various studies have been carried out on the RIS technology, which is capable of enabling the control of the signal propagation environment by network operators. However, when an RIS is used in its inherently passive structure, it appears to be only a supportive technology for communications, while suffering from a multiplicative path loss. Therefore, researchers have lately begun to focus on RIS hardware designs with minimal active elements to further boost the benefits of this technology. In this paper, we present a simple RIS hardware architecture including a single and variable gain amplifier for reflection amplification to confront the multiplicative path loss. The end-to-end signal model for communication systems assisted with the proposed amplifying RIS design is presented, together with an analysis focusing on the capacity maximization and theoretical bit error probability performance, which is corroborated by computer simulations. In addition, the major advantages of the proposed amplifying RIS design compared to its passive counterpart are discussed. It is shown that the proposed RIS-based wireless system significantly eliminates the double fading problem appearing in conventional passive RIS-assisted systems and improves the communication energy efficiency.
\end{abstract}


%
\IEEEpeerreviewmaketitle

\section{Introduction}

\IEEEPARstart{T}{he} number of mobile users trying to keep up with emerging technologies is increasing day by day. This situation has an impact on many things around the world, especially on communication networks. Recent generations of wireless communication have been developed to meet the high data demand arising from these technological developments \cite{Saad_6G_2020}. For these reasons, the fact that existing communication systems will start to become insufficient in the future, as it was before, will make the development of next generation of the wireless communication, 6G and beyond, inevitable. 6G focuses further enhancing data rate, security, consistency, and mobility, compared to its predecessors. In order to meet the future demands, numerous studies are being carried out towards 6G wireless networks. As a result, the use of advanced technologies such as extreme multiple-input multiple-output (MIMO) systems with power efficient front-ends \cite{shlezinger2021dynamic}, as well as Terahertz and millimeter-wave communications might play an important role in supplying these demands.

In recent years, reconfigurable intelligent surface (RIS)-empowered communication has become one of the most popular developments towards 6G networks \cite{basar2019wireless,energyeff,9482474,Marco2019,wu2020intelligent,9627818,CE_overview_2022}. Supporting high wireless channel capacity, expanding the signal coverage, reducing the bulkiness of the multiple antenna systems as well as the energy consumption, low-cost implementation and mitigating several negative effects of the wireless channel, such as multipath fading and the Doppler effect can be given as some of the main reasons why RISs has been investigated thoroughly\cite{yildirim2020modeling,Basar_Doppler,8910627}.

Unlike massive antenna arrays that include multiple radio frequency (RF) chains attached to the antennas, RISs include only a certain number of reflecting elements that reflect the incoming signals with a specific phase. This operation can be done by changing the electromagnetic properties of their reflecting elements, such as the reflection coefficient, with the help of a software-defined controller. In that way, the wireless propagation environment becomes configurable. The potential of RISs stems from the fact that they perform these operations while being almost passive because tiny reflecting elements are used instead of power-hungry phase shifters. Thus, energy efficient and low cost setups can be realized with the help of RISs.

Numerous studies have been conducted by combining RISs with many other techniques. Assisting a relay system with an RIS \cite{HybridRIS}, RIS-based index modulation\cite{basar2020reconfigurable}, RIS-enabled reflection modulation \cite{Miaowen2021,9405433}, implementation of non-orthogonal multiple access (NOMA) by using RISs \cite{9140006,khaleel2020novel}, and channel modeling in the presence of an RIS \cite{basar2020simris,basar2020indoor,kilinc2021physical} can be given as notable examples. These studies have succeeded with the help of RISs but have not been able to avoid the performance degradation caused by the double path loss problem when RIS is not positioned close to terminals \cite{basar2021present,dunna2020scattermimo}. Therefore, passive RISs cannot go beyond being a supportive technology for communication systems due to the double path loss problem. Some studies have been conducted to overcome the double path loss effect by combining active elements with RISs. In \cite{9053976}, a single receive RF chain that includes a low-noise amplifier (LNA) was used to enable baseband reception, which in turn, was exploited for channel estimation at the RIS side. Similarly, channel estimation using a hybrid RIS architecture has been performed in \cite{alexandropoulos2021hybrid} by deploying small numbers of receive RF chains at the RIS side. Another RIS-based channel sensing approach was studied in \cite{alamzadeh2021reconfigurable} by coupling some incident signals with a modified RIS architecture, including receive RF chains. Furthermore, \cite{shlezinger2021dynamic} analyzes dynamic metasurface antennas (DMAs) from an analog and digital signal processing perspective with less number of RF chains than the number of metamaterials. In \cite{zhang2021active}, a reflection-type amplifier has been proposed to be used by each reflecting element of an RIS, and the proposed active RIS element was designed, fabricated, and experimental results are provided. Another active reflecting element design has been proposed in \cite{9377648} such that the magnitude of the reflection coefficient of the reflecting elements is greater than unity. Connecting only a few elements of an RIS with RF chains and power amplifiers (PAs) has been investigated in \cite{nguyen2021hybrid} as a hybrid relay-reflecting intelligent surface. 

Against this background, this paper proposes an amplifying RIS design that incorporates a single PA and two RISs. According to our design, the signals received by elements of the first RIS are combined after getting their phase configured. Then, the combined signal is fed to the available PA, which feeds it to the second RIS that transmits it with a controllable phase configuration.

The proposed design works solely in the RF domain, similar to the waveguide-based approach in \cite{alamzadeh2021reconfigurable}, unlike full-duplex multi-antenna decode-and-forward (DF) relays, which perform down-conversion and baseband processing \cite{HybridRIS}. Therefore, our design looks more similar to a full-duplex, multi-antenna, and amplify-and-forward (AF) relay. However, they have some key differences. Relays usually perform linear processing techniques, such as maximum ratio combining (MRC), and realize power allocation optimization algorithms. Furthermore, they include bulky phase-shifter networks for transmit beamforming and are subject to loopback self-interference \cite{uysal2009cooperative}. Our design is simpler, does not require complex algorithms, and prevents loopback self-interference via spatial separation in the form of back to back placement of the different RISs. Note that simple self-interference cancellation techniques in the RF domain can be applied \cite{alexandropoulos2020full}. 

\begin{table}[]
\setlength{\extrarowheight}{0.15cm}
\caption{\textbf{Comparison of the proposed architecture with a passive RIS, a fully-connected active RIS, and a multi-antenna AF relay}}
\label{compare_table}
\resizebox{\columnwidth}{!}{%
\begin{tabular}{l|c|c|c|c|}
\cline{2-5}
                                                                                             & \textbf{\begin{tabular}[c]{@{}c@{}}Proposed \\ Design\end{tabular}} & \textbf{\begin{tabular}[c]{@{}c@{}}Passive \\ RIS\end{tabular}} & \textbf{\begin{tabular}[c]{@{}c@{}}Fully-\\ Connected \\ Active RIS\end{tabular}} & \textbf{AF Relay}                                                        \\ \hline
\multicolumn{1}{|l|}{\textbf{Complexity}}                                                          & Low                                                                 & Low                                                             & High                                                                             & High                                                                     \\ \hline
\multicolumn{1}{|l|}{\textbf{\begin{tabular}[c]{@{}l@{}}Power \\ Consumption\end{tabular}}}        & Low                                                                 & Low                                                             & High                                                                             & High                                                                     \\ \hline
\multicolumn{1}{|l|}{\textbf{\begin{tabular}[c]{@{}l@{}}Interference \\ Cancellation\end{tabular}}} & \begin{tabular}[c]{@{}c@{}}No \\ interference\end{tabular}                                                            & \begin{tabular}[c]{@{}c@{}}No \\ interference\end{tabular}      & \begin{tabular}[c]{@{}c@{}}No \\ interference\end{tabular}                       & \begin{tabular}[c]{@{}c@{}}Signal \\ processing \\ required\end{tabular} \\ \hline
\multicolumn{1}{|l|}{\textbf{Cost}}                                                                & Low                                                                 & Low                                                             & High                                                                             & High                                                                     \\ \hline
\multicolumn{1}{|l|}{\textbf{Performance}}                                                         & High                                                                & Low                                                             & High                                                                             & High                                                                     \\ \hline
\end{tabular}%
}
\end{table}
A detailed comparison of the proposed scheme with passive RISs, fully-connected active RIS and AF relay is presented in Table \ref{compare_table}. Considering fully-connected active designs, where each RIS element contains an active component, there are some drawbacks, such as power consumption and cost. These architectures may not be feasible for a large number of reflecting elements because they contain numerous active components. The proposed design has the key advantages of being energy efficient and low cost compared to other active RIS designs since it includes a single PA. Furthermore, the fully-connected active RIS architecture is more complex than the proposed system because the gain of each active element needs to be configured. In contrast, the gain of a single PA is adjusted in the proposed architecture. Although the fully-connected active RIS design is more powerful because of the amplifiers being connected to each reflecting element, the proposed scheme also benefits from the passive beamforming of the receiving and transmitting parts of the RIS. Compared to passive RISs, the proposed design enhances the system capacity in a more energy efficient manner. Moreover, as our system diminishes the effect of double path loss due to the amplification, it provides flexibility to place the RIS at any point between the transmitter (Tx) and the receiver (Rx). However, the proposed design is more complex because additional signal processing procedures are required, such as adjusting the gain of the PA and phase configurations of the RISs.

We note that a similar concept was considered in \cite{abari2017enabling}, assuming that the received signal is amplified and fed back as leakage to the input; however, it lacks a sufficient mathematical framework and does not present detailed modeling of the system operation. On the other hand, in this paper, we propose a novel amplifying RIS scheme with a solid mathematical framework and provide a unified model for the system operation. Moreover, we analyze the theoretical bit error probability (BEP), capacity, and energy efficiency (EE) of the proposed design and compare it with a benchmark passive RIS structure.

The main contributions of this study are summarized as follows:
\begin{itemize}
		\item A new and energy efficient amplifying RIS design is proposed. The end-to-end system model of the proposed amplifying RIS design is presented.
		
		\item This work demonstrates a cheaper, less complex, and more energy efficient architecture than fully connected active RIS designs and multi-antenna AF relays.
		\item The proposed design provides the flexibility to position the RIS at any location, with an increase in performance compared to passive RISs.
		\item The achievable rate and the EE of the system are examined under different configurations, and an optimization problem is addressed to maximize the system capacity. Moreover, the theoretical BEP is obtained under different conditions, while being also verified through computer simulations.
		\item The proposed active model is comprehensively compared with a passive RIS setup.
\end{itemize}

The rest of the paper is organized as follows. In Section II, we present the end-to-end system model of the proposed amplifying RIS design. Section III presents the performance analysis of the system by investigating the BEP with a theoretical approach. In Section IV, the EE analysis is performed by considering the characteristics of the PAs. In Section V, we exhibit our numerical results to evaluate the system performance. Finally, the paper is concluded in Section VI.

\section{System Model}
In this section, we propose an amplifying RIS-assisted system model. In addition, a conventional passive RIS system model is given as the benchmark to the proposed scheme.
\subsection{Amplifying RIS-assisted System}
\begin{figure}[!t]
	\begin{center}
		\includegraphics[width=0.9\columnwidth]{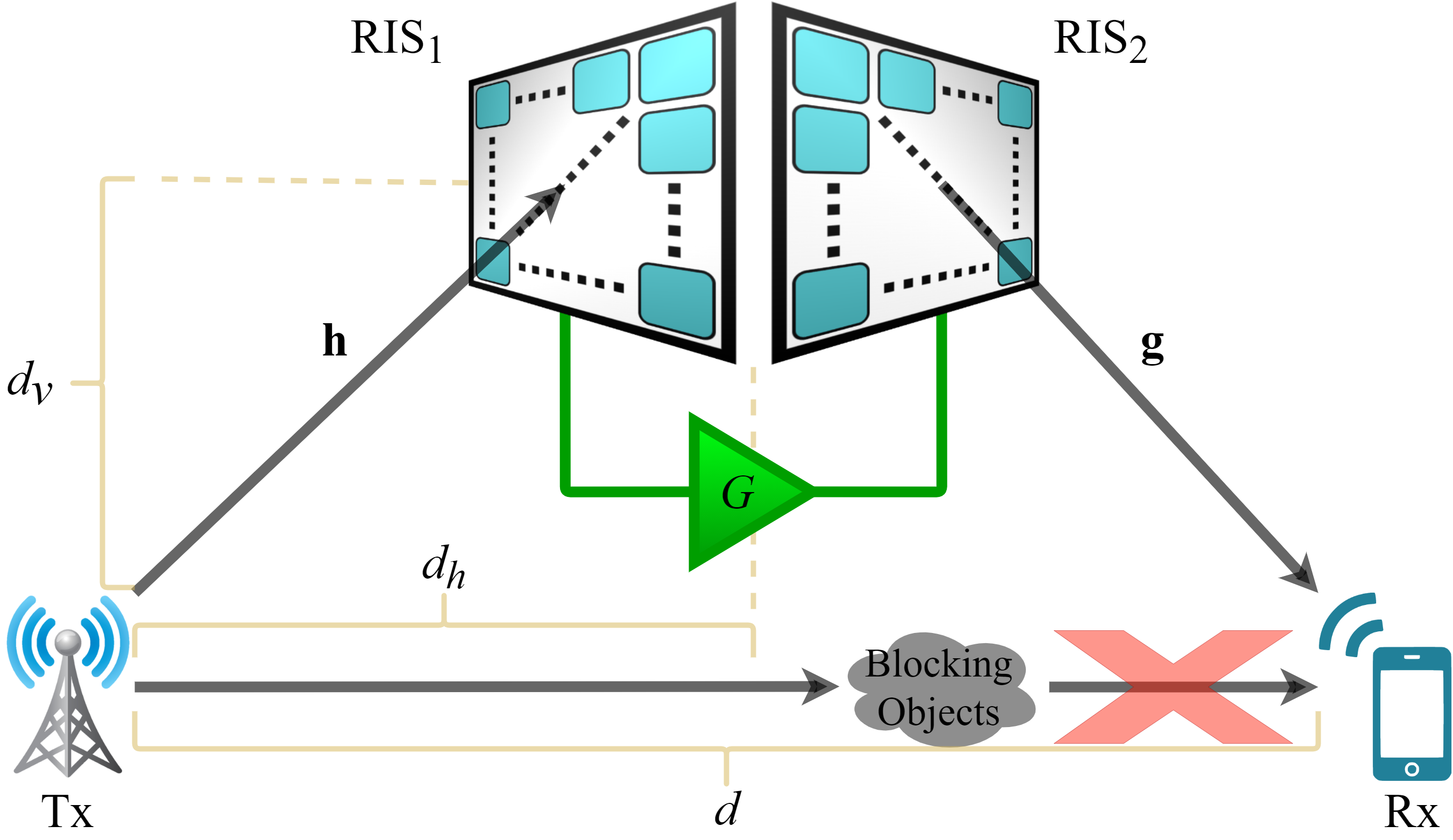}
		\caption{\textbf{Generic system model for the considered amplifying RIS-assisted scheme.}}
		\label{fig:Fig1}
	\end{center} \vspace*{-0.3cm}
\end{figure}

In this subsection, the considered single-input single-output (SISO) amplifying RIS-assisted system model is introduced. In this scenario, there are two passive RISs, denoted as RIS\textsubscript{1} and RIS\textsubscript{2}, that are connected with a PA, and both have $N$ number of reflecting elements while assuming there is no direct link between the Tx and Rx as shown in Fig. \ref{fig:Fig1}. The vertical and horizontal distances between the Tx and RISs, and the distance between the Tx and Rx are denoted as $d_v$, $d_h$ and $d$, respectively. All of the reflecting elements of RIS\textsubscript{1} receive signals from the Tx and these signals are combined by RIS\textsubscript{1} to be directed to the PA as a single signal. Then, the signal is amplified by the PA by taking into account that there is also some additional noise to be amplified, which is due to all of the reflecting elements of RIS\textsubscript{1}. Additionally, we consider the signal power at the output of the PA is limited. The system passes the resulting amplified signal through RIS\textsubscript{2}, such that the reflecting elements of RIS\textsubscript{2} share and transmit the amplified signal. One should consider that there might be leakage from the signal transmitted by RIS\textsubscript{2} to the signal received by RIS\textsubscript{1}. However, this issue can be resolved by properly positioning the RISs, since an RIS is an one-sided surface. In our design, RIS\textsubscript{1} and RIS\textsubscript{2} are placed back-to-back and separately as shown in Fig. \ref{fig:Fig1}, so that the signals directed by RIS\textsubscript{2} do not contaminate the signal captured by RIS\textsubscript{1}.

In light of these, the received complex baseband signal can be expressed as
\begingroup
\setlength\abovedisplayskip{10pt}
\setlength\belowdisplayskip{10pt}
\begin{equation} \label{eq:1}
y=\sqrt{\frac{G}{N}}\left(\boldsymbol\phiup^\mathrm{T}\mathbf{h}\sqrt{P_t}s+\sqrt{F}n_\text{tot}\right)\boldsymbol\thetaup^\mathrm{T}\mathbf{g}+n_\text{rx},
\end{equation}
\endgroup
where $s$, $y$, $P_t$, $G$, $F$, $n_\text{tot}$, and $n_\text{rx}$, correspond to the transmitted and received signals, transmit power at the Tx, the gain and noise figure of the PA, the total amount of noise at the  input of the PA, and the noise sample at the Rx, respectively. $\boldsymbol\phiup\in \mathbb{C}^{N\times 1}$ and $\boldsymbol\thetaup\in \mathbb{C}^{N\times 1}$ are the reflecting element phase shift vectors of RIS\textsubscript{1} and RIS\textsubscript{2}, respectively, where $\boldsymbol\phiup=\left[\phi_1,\phi_2,\dots,\phi_N\right]^\mathrm{T}$, $\boldsymbol\thetaup=\left[\theta_1,\theta_2,\dots,\theta_N\right]^\mathrm{T}$, and $\phi_i$ and $\theta_i$ denote the phase shifts of the $i$th reflecting element of RIS\textsubscript{1} and RIS\textsubscript{2} for $i=1,\dots,N$, respectively.
$\mathbf{h}\in \mathbb{C}^{N\times 1}$ is the channel between the Tx and RIS\textsubscript{1}, where $\mathbf{h}=\left[h_1,h_2,\dots,h_N\right]^\mathrm{T}$, and $h_i$ denotes the channel coefficient for Tx and $i$th reflecting element of RIS\textsubscript{1} for $i=1,\dots,N$. $\mathbf{g}\in \mathbb{C}^{N\times 1}$ is the channel between RIS\textsubscript{2} and the Rx, where $\mathbf{g}=\left[g_1,g_2,\dots,g_N\right]^\mathrm{T}$, and $g_i$ represents the channel coefficient for $i$th reflecting element of RIS\textsubscript{2} and the Rx for $i=1,\dots,N$. Here, $\boldsymbol\phiup^\mathrm{T}\mathbf{h}\sqrt{P_t}s$ stands for the collected signal received by RIS\textsubscript{1}. This signal is scaled by $\sqrt{G}$ with the help of the PA and divided by $\sqrt{N}$ due to the phase shift network of RIS\textsubscript{2} that is a kind of power-division circuit, so the total power is distributed among all of the reflecting elements of RIS\textsubscript{2}. After that, the amplified and distributed signal is steered to the Rx with the help of RIS\textsubscript{2}. A similar scenario is also valid for $n_\text{tot}$, so that it is multiplied with $\sqrt{F}$ and $\sqrt{G}$, then distributed among the reflecting elements of RIS\textsubscript{2}. Here, the noise at RIS\textsubscript{2} is ignored, where the noise at RIS\textsubscript{1} is not because it is subject to amplification.

The channels exhibit either Rayleigh or Rician fading depending on the line-of-sight (LOS) probability, $p_\text{LOS}$, which is a function of the distance. The channel coefficients are modeled as
\begin{equation} \label{eq:2}
    h_i =  \sqrt{\frac{1}{\lambda^h}} \left(\sqrt{\frac{K_1}{K_1+1}} h^L_i + \sqrt{\frac{1}{K_1+1}}h^{NL}_i\right),
\end{equation}
\begin{equation} \label{eq:3}
    g_i =  \sqrt{\frac{1}{\lambda^g}} \left(\sqrt{\frac{K_2}{K_2+1}} g^L_i + \sqrt{\frac{1}{K_2+1}}g^{NL}_i\right),
\end{equation}
where $K_1$, $K_2$, $\lambda^h$, $\lambda^g$, $h^L_i$, $g^L_i$, $h^{NL}_i$ and $g^{NL}_i$ denote the Rician factors for Tx-RIS\textsubscript{1} and RIS\textsubscript{2}-Rx links, the path loss, LOS and non-line-of-sight (NLOS) components for the channels $\mathbf{h}$ and $\mathbf{g}$, respectively, $h^{NL}_i,g^{NL}_i \sim\mathcal{CN}(0,1)$ for $i=1,\dots,N$, and $\mathcal{CN}(0,1)$ stands for the complex Normal distribution with zero mean and unit variance. If the channels do not include a LOS component, which mostly refers to Rayleigh fading, we consider $K_1=K_2=0$. Furthermore, the path loss components $\lambda^h$ and $\lambda^g$ depend on having NLOS or LOS links, and calculated as follows by considering the Indoor Hotspot (InH) environment in \cite{3GPP_5G}:
\begingroup\makeatletter\def\f@size{9}\check@mathfonts
\begin{equation} \label{eq:4}
    \lambda_{LOS} \text{[dB]}= 32.4+17.3\log_{10}(d_n)+20\log_{10}(f_c),
\end{equation}
\endgroup
\begingroup\makeatletter\def\f@size{9}\check@mathfonts
\begin{equation} \label{eq:5}
    \lambda_{NLOS} \text{[dB]}= \max\left(\lambda_{LOS}, 32.4+31.9\log_{10}(d_n)+20\log_{10}(f_c)\right),
\end{equation}
\endgroup
where $d_n \in \{d_1,d_2\}$ depends on the channel, and $d_1=\sqrt{d_v^2+d_h^2}$, $d_2=\sqrt{d_v^2+(d-d_h)^2}$, and $f_c$ are denoted as the distances between the Tx-RIS\textsubscript{1}, RIS\textsubscript{2}-Rx, and carrier frequency in GHz, respectively. The path loss components are the same for each $h_i$ and $g_i$ because the RIS is located in the far-field of the Tx and Rx. Furthermore, $p_\text{LOS}$ for the InH environment is given as \cite{3GPP_5G}

\begin{equation}\label{eq:6}
p_\text{LOS}=\begin{cases}
1, & d_n \le 5, \\
e^{-\left(\frac{d_n -5}{70.8} \right), } & 5<d_n \le 49, \\
0.54e^{-\left(\frac{d_n -49}{211.7}\right), } & 49<d_n.
\end{cases} 
\end{equation}

To determine the signal-to-noise ratio (SNR) of the amplifying RIS-assisted sytem, \eqref{eq:1} is expanded as
\begin{equation} \label{eq:7}
y=\sqrt{\frac{GP_t}{N}}\left(\boldsymbol\phiup^\mathrm{T}\mathbf{h}\right)\left(\boldsymbol\thetaup^\mathrm{T}\mathbf{g}\right)s+\sqrt{\frac{GF}{N}}\left(\boldsymbol\thetaup^\mathrm{T}\mathbf{g}\right)n_\text{tot}+n_\text{rx}.
\end{equation}
From the received signal model given in \eqref{eq:7}, the SNR of the proposed system as follows:
\begingroup
\setlength\abovedisplayskip{10pt}
\setlength\belowdisplayskip{10pt}
\begin{equation} \label{eq:8}
\gamma_\text{act}=\dfrac{P_t\left|\sqrt{\dfrac{G}{N}}(\boldsymbol\phiup^\mathrm{T}\mathbf{h})(\boldsymbol\thetaup^\mathrm{T}\mathbf{g})\right|^2}{\left|\sqrt{\dfrac{GF}{N}}\boldsymbol\thetaup^\mathrm{T}\mathbf{g}\right|^2\sigma^2_\text{tot}+\sigma^2_\text{rx}},
\end{equation}
\endgroup
where $\sigma^2_\text{tot}$ and $\sigma^2_\text{rx}$ are the noise powers at the input of the PA and Rx, respectively. $\gamma_\text{act}$ is the instantaneous received SNR of the amplifying RIS-assisted system and the achievable rate of the system is expressed as $R_\text{act}=\log_{2}\left(1+\gamma_\text{act}\right)$.

\begin{figure}[!t]
	\begin{center}
		\includegraphics[width=0.9\columnwidth]{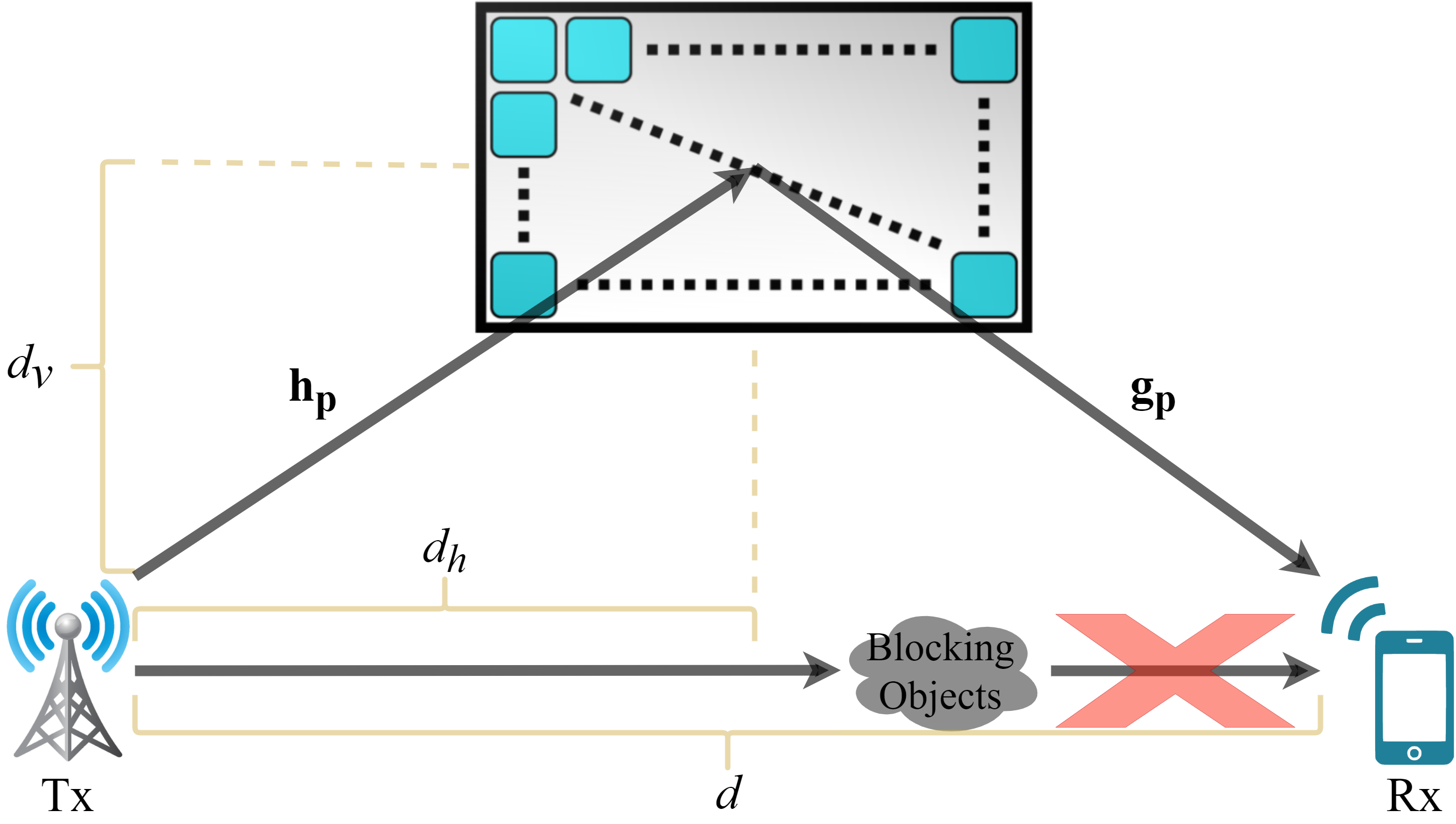}
		\caption{\textbf{Generic system model for a passive RIS-assisted scheme.}}
		\label{fig:Fig2}
	\end{center} \vspace*{-0.3cm}
\end{figure}
\subsection{Passive RIS-assisted System}
This section is based on a passive RIS-assisted model shown in Fig. \ref{fig:Fig2}. In this scenario, only one RIS is used and it contains $2N$ reflecting elements such that it can be considered as a fair benchmark to the amplifying RIS-assisted model. The signal model of the passive system is specified as
\begin{equation} \label{eq:12}
    y=\mathbf{g_p}^\mathrm{T}\mathbf{\Psi} \mathbf{h_p}\sqrt{P_t}s+n_\text{rx},\vspace{0.2cm}
\end{equation}
where $\mathbf{\Psi} \in \mathbb{C}^{2N\times 2N}$ is the reflecting element phase shift matrix of the RIS and $\mathbf{\Psi}=\mathrm{diag} \left( \begin{bmatrix} \psi_1,\psi_2,\dots,\psi_{2N} \end{bmatrix}   \right)$ with $\psi_i$ representing the phase shift for the $i$th element of the RIS for $i=1,\dots,2N$. $\mathbf{h_p}\in \mathbb{C}^{2N\times 1}$ is the channel between the Tx and RIS, while $\mathbf{g_p}\in \mathbb{C}^{2N\times 1}$ stands for the channel between the RIS and Rx. They follow either Rician or Rayleigh fading depending on the distances to the Tx and Rx similar to the channels in the active model. Here, SNR of the passive model $\gamma_\text{pas}$ is determined as
\begin{equation} \label{eq:13}
 \gamma_\text{pas}=\dfrac{P_t\left|\mathbf{g_p} ^\mathrm{T} \mathbf{\Psi} \mathbf{h_p}\right|^2}{\sigma^2_\text{rx}},
\end{equation}
where the achievable rate for the passive model is $R_\text{pas}=\log_{2}\left(1+\gamma_\text{pas}\right)$.

\section{Performance Analysis}
In this section, we analyze the maximization of system capacity and the distribution of the received SNR for the amplifying RIS-assisted system, and present our theoretical BEP calculations accordingly. In order to maximize the system capacity, the received SNR for the amplifying RIS-assisted system should be maximized by finding optimum values of the phase shift vectors of RIS\textsubscript{1}, RIS\textsubscript{2} and the gain of the amplifier $G$. The corresponding optimization problem is formulated as follows:
\begin{align}
        \gamma_\text{act}=&\max_{G,\boldsymbol\phiup,\boldsymbol\thetaup}\quad \dfrac{P_t\left|\sqrt{\dfrac{G}{N}}(\boldsymbol\phiup^\mathrm{T}\mathbf{h})(\boldsymbol\thetaup^\mathrm{T}\mathbf{g})\right|^2}{\left|\sqrt{\dfrac{GF}{N}}\boldsymbol\thetaup^\mathrm{T}\mathbf{g}\right|^2\sigma^2_\text{tot}+\sigma^2_\text{rx}} \nonumber \\
        &\hspace*{0.3cm}\text{s.t.}\hspace*{0.45cm} \left|\phi_i\right|= \left|\theta_i\right|=1 , \quad  i\in\{1,2,\dots,N\}, \nonumber \\ &\hspace*{1.24 cm}
        GP_t\left|\boldsymbol\phiup^\mathrm{T}\mathbf{h}\right|^2\leq P_\text{max}, \nonumber \\ &\hspace*{1.24 cm} G \leq G_\text{max},
\end{align}
where $G_\text{max}$ and $P_\text{max}$ stand for the maximum gain of the amplifier and the maximum output power of the amplifier, respectively. Here, the optimal value of $G$ depends on the phase shift matrix of RIS\textsubscript{1} as well. Therefore, the optimal solution to the phases of the reflecting elements should be determined first. The signals should be constructively combined at the PA and Rx to increase the received signal power. This can be done by eliminating the phases of the channels $\mathbf{h}$ and $\mathbf{g}$. The optimal solutions to reflecting element phases for RIS\textsubscript{1} and RIS\textsubscript{2} follow a constructive combining strategy \cite{8811733} by adjusting the phase shifts of the reflecting elements as:
\begin{table*}[t]
\setlength{\extrarowheight}{8pt}
\centering
\caption{\textbf{The shape ($k$) and scale ($\nu$) parameters of the fitted Gamma distribution}}
\label{Fittable}
\begin{tabular}{cclccccccccccc}
\hline
                                                & \multicolumn{2}{c}{}                     & $N$                                         & $P_t\text{(dBm)}$: & -10      & -5       & 0        & 5        & 10       & 15       & 20       & 25       & 30       \\    \hline \hline 
\multicolumn{1}{c|}{\multirow{8}{*}{\rotatebox[origin=c]{90}{$P_\text{{max}}\text{(dBm)}\quad\quad\quad\quad$}}} & \multicolumn{2}{c|}{\multirow{4}{*}{$\vspace{-0.6cm} 10$}} & \multicolumn{1}{c|}{\multirow{2}{*}{$\vspace{-0.3cm}64$}}  & $k$        & 44.8922  & 44.7180  & 44.7905  & 44.7109  & 44.8358  & 44.8963  & 48.0934  & 58.6428  & 58.7049  \\ [.5ex]
\multicolumn{1}{c|}{}                           & \multicolumn{2}{c|}{}                    & \multicolumn{1}{c|}{}                     & $\nu$        & 0.000405 & 0.001287 & 0.004063 & 0.012868 & 0.040595 & 0.128166 & 0.375257 & 0.329909 & 0.329483  \\ [1.5ex] \cline{4-14} 
\multicolumn{1}{c|}{}                           & \multicolumn{2}{c|}{}                    & \multicolumn{1}{c|}{\multirow{2}{*}{$\vspace{-0.3cm}256$}} & $k$        & 178.8281 & 178.8481 & 179.1008 & 178.2996 & 233.7027 & 234.8395 & 234.5761 & 233.9432 & 234.3112 \\ [.5ex]
\multicolumn{1}{c|}{}                           & \multicolumn{2}{c|}{}                    & \multicolumn{1}{c|}{}                     & $\nu$        & 0.006486 & 0.020508 & 0.064762 & 0.205720 & 0.330019 & 0.328426 & 0.328834 & 0.329725 & 0.329212 \\ [1.5ex] \cline{2-14} 
\multicolumn{1}{c|}{}                           & \multicolumn{2}{c|}{\multirow{4}{*}{$\vspace{-0.6cm}20$}} & \multicolumn{1}{c|}{\multirow{2}{*}{$\vspace{-0.3cm}64$}}  & $k$        & 44.8284  & 44.7199  & 44.7593  & 44.84389 & 44.8633  & 44.7232  & 44.7599  & 44.7586  & 47.9274  \\ [.5ex]
\multicolumn{1}{c|}{}                           & \multicolumn{2}{c|}{}                    & \multicolumn{1}{c|}{}                     & $\nu$        & 0.000406 & 0.001286 & 0.004066 & 0.012835 & 0.040551 & 0.128680 & 0.406483 & 1.285651 & 3.765564 \\ [1.5ex] \cline{4-14} 
\multicolumn{1}{c|}{}                           & \multicolumn{2}{c|}{}                    & \multicolumn{1}{c|}{\multirow{2}{*}{$\vspace{-0.3cm}256$}} & $k$        & 178.2811 & 178.9068 & 178.4629 & 178.8242 & 178.8931 & 178.6688 & 234.2556 & 234.8709 & 233.5700 \\ [.5ex]
\multicolumn{1}{c|}{}                           & \multicolumn{2}{c|}{}                    & \multicolumn{1}{c|}{}                     & $\nu$        & 0.006505 & 0.020503 & 0.064994 & 0.205099 & 0.648377 & 2.052960 & 3.292076 & 3.283723 & 3.302386 \\ [1.5ex] \hline
\end{tabular}
\end{table*}
\begingroup
\setlength\abovedisplayskip{7pt}
\setlength\belowdisplayskip{10pt}
\begin{align} \label{eq:phase1} 
    \phi_i=e^{-j\angle h_i}, \\ \label{eq:phase2}
    \theta_i=e^{-j\angle g_i},
\end{align}
\endgroup
where $\angle\cdot$ denotes the phase of a complex term. Next, we should specify the value for an optimal $G$ by fixing the optimal value of $\boldsymbol{\phiup}$ as in \eqref{eq:phase1}. It is mentioned earlier that there is a maximum power level for the amplified signal. Therefore, an optimal gain value should be determined so that the power of the input signal to the amplifier, $P_\text{in}=P_t\left(\sum\nolimits_{i=1}^{N}\left|h_i\right|\right)^2$, should not exceed the maximum output power when it is amplified. Thus, the equation below should be satisfied to obtain the optimal value of $G$;\vspace*{-0.2 cm}
\begingroup
\setlength\abovedisplayskip{15pt}
\setlength\belowdisplayskip{12pt}
\begin{align} 
    \overline{G}_\text{opt}P_t\left(\sum\limits_{i=1}^{N}\left|h_i\right|\right)^2=P_\text{max},
\end{align} 
\endgroup
where $\overline{G}_\text{opt}$ stands for the optimal gain value without the limitation of $G_\text{max}$. Considering that $\overline{G}_\text{opt}$ can exceed $G_\text{max}$, we define the optimal gain as:
\begingroup
\setlength\abovedisplayskip{12pt}
\setlength\belowdisplayskip{15pt}
\begin{align} \label{eq:Gopt}
    G_\text{opt} = \min\left(\overline{G}_\text{opt},G_\text{max}\right).
\end{align}
\endgroup
By substituting $G_\text{opt}$ in \eqref{eq:Gopt} to \eqref{eq:8} for $G$ and arranging the reflecting element phases of $\boldsymbol{\phiup}$ and $\boldsymbol{\thetaup}$ as in \eqref{eq:phase1} and \eqref{eq:phase2}, respectively, the maximized received SNR for the amplifying RIS-assisted system can be written as
\begingroup
\setlength\abovedisplayskip{12pt}
\setlength\belowdisplayskip{12pt}
\begin{align} \label{eq:gamma_active}
    \gamma_\text{act}&=\dfrac{P_t\dfrac{P_\text{max}}{N} A^2B^2}{\dfrac{P_\text{max}F}{N}B^2\sigma_\text{tot}^2+P_tA^2\sigma_\text{rx}^2},
\end{align} 
\endgroup
where $A=\sum\nolimits_{i=1}^{N}\left|h_i\right|$ and $B=\sum\nolimits_{i=1}^{N}\left|g_i\right|$. Here, one can easily observe that the maximized received SNR increases proportional to $N$, $P_t$, and $P_\text{max}$. However, there are also other constraints that can limit the effect of the RIS size and the Tx power, such as $P_\text{max}$ and $G_\text{max}$. The effects of these constraints are further investigated in Section V. Moreover, the amplifying RIS model experiences multiplicative path loss just like the passive RIS since \eqref{eq:gamma_active} includes the multiplication of $A$ and $B$. Nevertheless, the gain factor significantly compensates this double path loss effect.

Probability analysis of the maximized SNR in \eqref{eq:gamma_active} gives us useful insights. The terms $A$ and $B$ converge to Gaussian distributed random variables due to the Central Limit Theorem (CLT) for sufficiently large $N$, where $\left|h_i\right|$ and $\left|g_i\right|$ are Rayleigh or Rician distributed random variables. Thus, $A^2$ and $B^2$ follow non-central chi-square distribution with one degree of freedom. The denominator and numerator of the SNR term in \eqref{eq:gamma_active} are correlated and follow Gamma distribution due to the combination of weighted non-central chi-square random variables. The received SNR term in \eqref{eq:gamma_active} becomes the ratio of two correlated Gamma random variables and therefore, subject to the Gamma distribution \cite{flueck1973moments}. To the best of the authors' knowledge, although there are a number of articles in the literature examining the ratio of two Gamma random variables and ratio of correlated chi-square random variables, no studies pointing to this kind of distribution has been found\cite{joarder2009moments,lee1979distribution,provost1994exact}. Therefore, using Open Distribution Fitter app (dfittool), we can find the Gamma parameters of the received SNR distribution on MATLAB and prove that it follows the Gamma distribution.

\begin{figure}[t]
	\begin{center}
		\includegraphics[width=1\columnwidth]{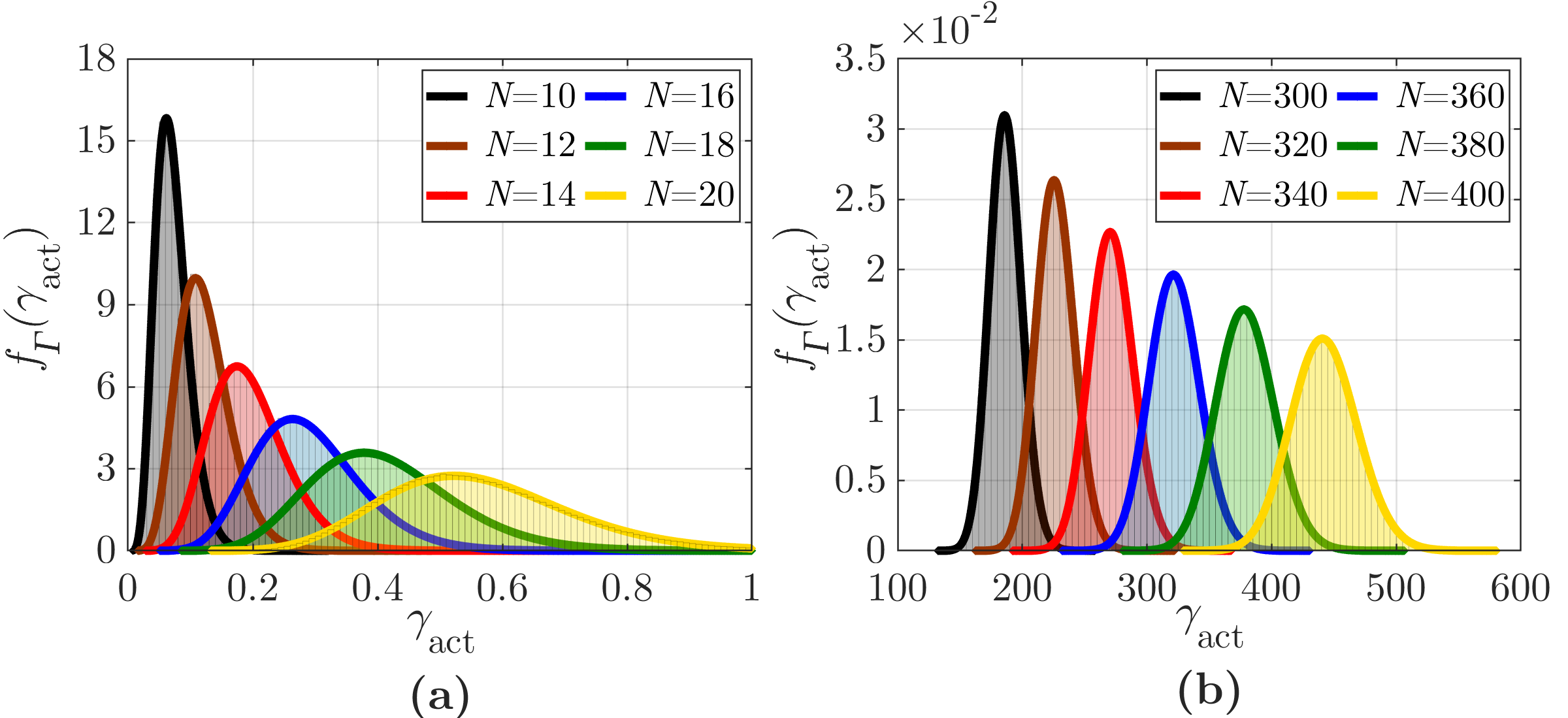}
		\caption{\textbf{The PDF of $\gamma_\text{act}$ and Gamma distribution fits for (a) \boldmath$N=10,\dots,20$ and (b) \boldmath$N=300,\dots,400$.}}
		\label{fig:gammas}
	\end{center} \vspace*{-0.5cm}
\end{figure}

The distribution of $\gamma_{\text{act}}$ is investigated for different $N$ values as seen in Fig. \ref{fig:gammas}. Histograms represent the distributions of $\gamma_{\text{act}}$ and are obtained by performing Monte Carlo simulations. The overlapping lines stand for the fitted Gamma distributions. The Gamma distribution fits exactly to the probability density function (PDF) of $\gamma_\text{act}$. Thus, the PDF of $\gamma_{\text{act}}$ can be written as\cite{lee1979distribution}
\begin{align}\label{eq:Mgf}
f_\Gamma(\gamma_{\text{act}}) = \frac{{\gamma_{\text{act}}^{k-1}e^{-\gamma_{\text{act}}/v}}}{v^k\Gamma(k)},
\end{align}
where $\Gamma(\cdot)$, $k$, and $\nu$ represent the Gamma function, shape, and scale parameters, respectively. The shape and scale parameters for the fitted Gamma distribution are provided in Table \ref{Fittable} for various system specifications\footnote{Interested readers can produce arbitrary results as well by using the source code provided: https://github.com/recepakiftasci/Amplifying-RIS}. The parameters of the Gamma distribution vary for different specifications such as $N$, $P_\text{max}$ and $P_t$. For instance, the standard deviation of the distribution increases with $N$ as shown in Fig. \ref{fig:gammas}.

Theoretical symbol error probability (SEP) is calculated by using the moment generation function (MGF) of the Gamma distribution as follows \cite{arslan2021over}: 
\begin{align}\label{eq:Mgf}
M_{\gamma_\text{act}}(s)& = (1-\nu s)^{-k},\hspace{.2cm} \text{for} \hspace{.2cm}   s<\cfrac{1}{\nu}.
\end{align}
The average SEP for $M$-ary phase-shift keying ($M$-PSK) signaling by using the MGF in  \eqref{eq:Mgf} is given as follows:
\begingroup
\setlength\abovedisplayskip{15pt}
\setlength\belowdisplayskip{15pt}
\begin{align} \label{eq:SEP}
P_s& = \cfrac{1}{\pi}\bigintss_{0}^{(M-1)\pi/M}M_{\gamma_\text{act}} \left( \cfrac{-\text{sin}^2(\pi/M)}{\text{sin}^2x}\right)dx.
\end{align}
\endgroup
We can numerically calculate the SEP in \eqref{eq:SEP} by using the Gamma distribution parameters and obtain BEP as $P_e \approx P_s/\log_2M$. Accordingly, the BEP for binary phase-shift keying (BPSK) simplifies to 
\begingroup
\setlength\abovedisplayskip{15pt}
\setlength\belowdisplayskip{15pt}
\begin{align}
P_e& = \cfrac{1}{\pi}\bigintss_{0}^{\pi/2}\left(1+\cfrac{\nu}{\text{sin}^2x}\right)^{-k}dx.
\end{align}
\endgroup

\section{Total Power Consumption Model and Energy Efficiency Analysis}

In this section, we present power consumption models for the proposed amplifying and passive RIS designs, and we analyze and compare the EE for both systems. The main power consuming elements can be listed as transmitter, receiver, amplifiers and RIS elements. 
There are two PAs in the system, one at the Tx and one between the RISs. We assume ideal PAs whose power efficiency is given as \cite{amplifier2}
\begingroup
\abovedisplayskip=15pt
\belowdisplayskip=15pt
\begin{align} \label{eq:power_c}
    \frac{P_\text{out}}{P_\text{amp}}=\eta_\text{max}\left(\frac{P_{\text{out}}}{P_{\text{max}}}\right)^{\varepsilon},
\end{align}
\endgroup
where $P_{\text{amp}}$ and $P_{\text{out}}$ correspond to power consumed by the amplifier and the output power of the amplifier, respectively. The maximum output power of the amplifier is set to $P_\text{max}$ to ensure that it operates in the linear region. Here, $\eta_{\text{max}}\in(0,1]$ is the maximum efficiency of the amplifier and $\varepsilon$ is a parameter that depends on the amplifier class. We assume $\varepsilon=0.5$ for more accurate modeling as in \cite{amplifier1}. The power consumed by the PA can be obtained by reorganizing \eqref{eq:power_c} as
\begingroup
\abovedisplayskip=10pt
\belowdisplayskip=15pt
\begin{align} \label{eq:pamp}
    P_{\text{amp}}=\frac{1}{\eta_{\text{max}}}\sqrt{P_{\text{out}}P_{\text{max}}}.
\end{align}
\endgroup
The phase shift of each RIS element is arranged by programmable electronic circuits that consume power as well. The power consumption of the RIS depends on the phase resolution of RIS elements \cite{energyeffD2D} and modeled as
\begingroup
\abovedisplayskip=10pt
\belowdisplayskip=10pt
\begin{align}
    P_\text{RIS} = N P_n(b),
\end{align}
\endgroup
where $P_n (b)$ is the power consumption of each RIS element which is a function of bit-resolution. We consider $6$-bit phase resolution for each RIS element, which consumes $7.8$ mW power according to \cite{energyeff}.
The total power consumption of the amplifying RIS-assisted system is expressed as follows:
\begingroup
\abovedisplayskip=15pt
\belowdisplayskip=15pt
\begin{align}
    P_\text{tot}^\text{act}=\alpha P_t + P_\text{Tx} + P_\text{Rx} + NP_{n}(b) + \beta\sqrt{P_\text{out}P_\text{max}} ,
\end{align}
\endgroup
where $\alpha=\omega_\text{max}^{-1}$ and $\beta=\eta_\text{max}^{-1}$ with $\omega_\text{max}$  and $\eta_\text{max}$ represent the maximum efficiency of the transmit PA and the PA between the RISs, respectively. We assume that $P_t=P_\text{max}$ for the transmit PA. Here, $P_\text{Tx}$ and $P_\text{Rx}$ are the hardware dissipated static powers at Tx and Rx, respectively and therefore they are constant and do not depend on the system parameters. The values for the power consumption model parameters are given in Table \ref{powerpar} as stated in \cite{energyeff}. Likewise, the power consumption of the passive RIS-assisted system can be written as follows only by omitting the power consumed by the PA between the RISs:
\begin{table}[t]
\setlength{\extrarowheight}{7pt}
\centering
\caption{\textbf{Power consumption model parameters}}
\label{powerpar}
\begin{tabular}{ | c | c | }
\hline
\centering
\textbf{Parameter} & \textbf{Value}    \\ [.5ex] \hline 
$\alpha$ & 1.2    \\ [.5ex] \hline
$\beta$  & 1.2    \\ [.5ex]\hline
$P_n(b)$  & 7.8 mW \\ [.5ex]\hline
$P_\text{Tx}$  & 9 dBW  \\ [.5ex]\hline
$P_\text{Rx}$    & 10 dBm \\ [.5ex]\hline
\end{tabular}
\end{table}
\begingroup
\abovedisplayskip=15pt
\belowdisplayskip=15pt
\begin{align}
    P_\text{tot}^\text{pas}=\alpha P_t + P_\text{Tx} + P_\text{Rx} + NP_{n}(b),
\end{align}
\endgroup
The bit-per-joule energy efficiency $\left(\eta_\text{EE}\right)$ of a system can be expressed as

\begingroup
\abovedisplayskip5pt
\belowdisplayskip=15pt
\begin{align}
   \eta_\text{EE}=\frac{R_i BW}{P_\text{tot}^i},
\end{align}
\endgroup
where $BW$ is the communication bandwidth, $R_i$ is the achievable rate, and $P_\text{tot}^i$ is the total consumed power by the system where $i\in\{\text{act},\text{pas}\}$. The amplifying RIS-assisted system consumes more power but provides higher capacity in return. On the other hand, passive RIS-assisted system is less power consuming and provides less capacity due to its fully passive nature. The EE analysis is an important criterion determining which system performs better in terms of energy consumption. In this study, we compute EE for different systems through computer simulations without particularly focusing on the optimization of the EE.

\section{Numerical Results}

In this section, we provide numerical results for both the amplifying and passive RIS models to evaluate and compare the performances of them under several configurations. Achievable rate, bit error rate (BER), power consumption, and EE have been presented by computer simulations and discussed in detail. We consider the systems as shown in Figs. \ref{fig:Fig1} and \ref{fig:Fig2} with parameters $d_v=5$ m, $d_h=5$ m, $d=50$ m, $P_t=30$ dBm, $P_\text{max}=30$ dBm, $f_c=28$ GHz, $BW=180$ kHz, $F=5$ dB, $K_1=K_2=5$, $N=128$, $n_\text{rx}=n_\text{tot}=-100$ dBm, and $G_\text{max}=30$ dB, unless specified otherwise\cite{HMC906A,PA_sample,HMC7054}. Achievable rate and EE simulations are performed with $10^6$ iterations. 
\begin{figure}[t]
	\begin{center}
		\centerline{\includegraphics[width=1.1\columnwidth]{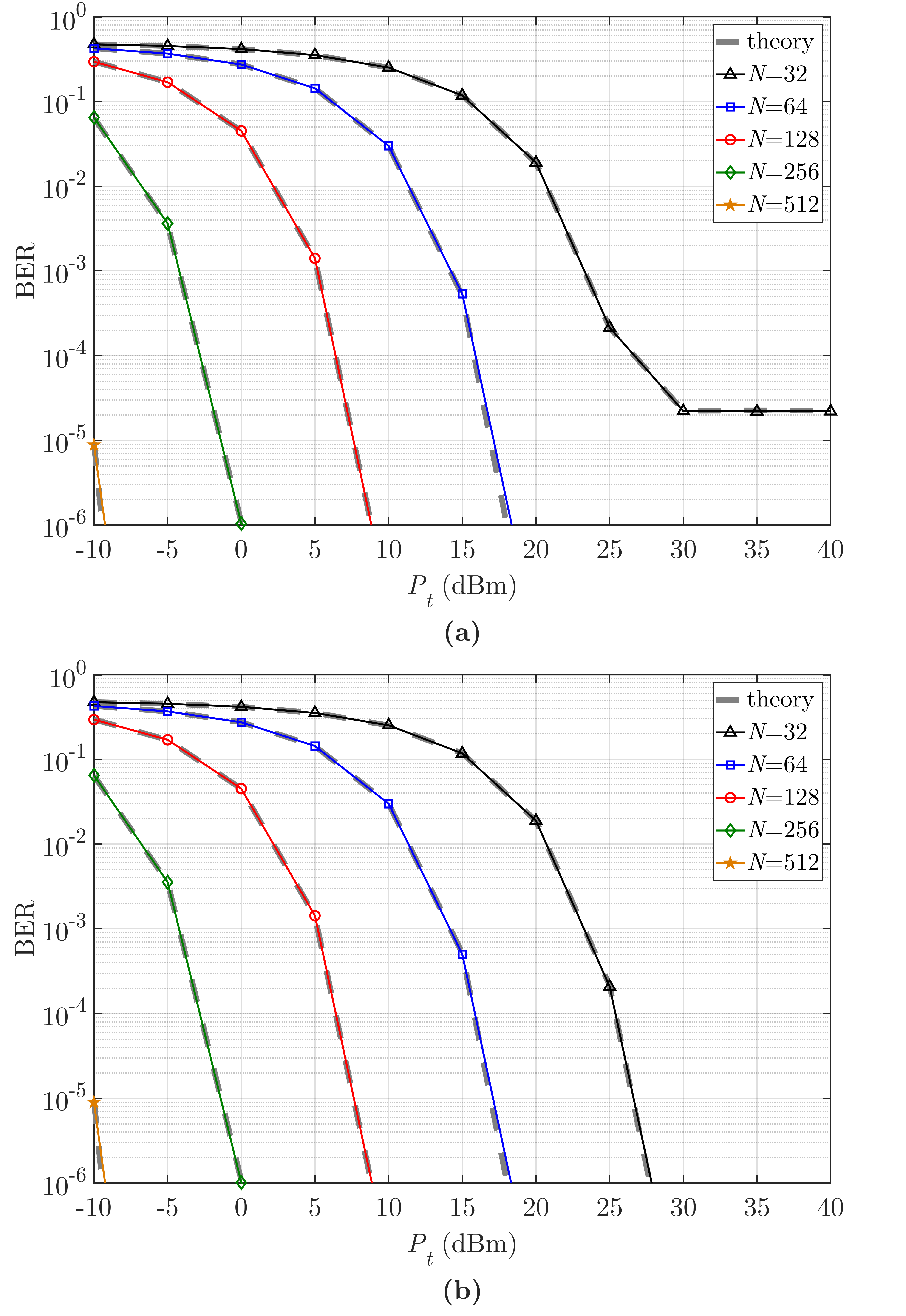}}
		\caption{\textbf{BER results for (a) \boldmath$P_\text{max}=10$ dBm and (b) \boldmath$P_\text{max}=20$ dBm.}}
		\label{fig:Fig3}
	\end{center} \vspace*{-0.5cm}
\end{figure}

\subsection{Performance Evaluation}

In this subsection, numerical results for BER and achievable rate are presented. For the BER analysis, two different setups with $P_\text{max}=10$ dBm and $P_\text{max}=20$ dBm are considered. Fig. \ref{fig:Fig3}(a) shows the BER performance when $P_\text{max}=10$ dBm, while Fig. \ref{fig:Fig3}(b) exhibits the results for $P_\text{max}=20$ dBm. Here, the corresponding theoretical values are given, as well. From Fig. \ref{fig:Fig3}(a), one can easily observe that an error floor occurs after $P_t=25$ dBm for $N=32$ case. On the other hand, an error floor is not observed in Fig. \ref{fig:Fig3}(b). This is because of the limitation of $P_\text{out}$ when the input signal becomes too strong such that the amplifier cannot stay in the linear region if it continues to enhance the signal with the current gain. In this case, when $P_\text{out}$ is fixed to $P_\text{max}$, further increment in $P_t$ does not lead a better error performance after a certain point, and thus we observe an error floor. 
On the other hand, in Figs. \ref{fig:Fig3}(a) and (b)  we do not observe any error floor when $P_\text{out}$ is smaller than $P_\text{max}$. Furthermore, the better BER performance is achieved with larger $N$ and higher $P_t$ in both cases as long as $P_\text{out}$ does not reach $P_\text{max}$. Based on these investigations, we can conclude that it may be beneficial to keep $P_\text{max}$ high enough to prevent any error floor. In addition, it should be emphasized that using a larger RIS can further strengthen the received signal and an error floor appears for lower $P_t$ values due to the limitation of $P_\text{out}$.
\begin{figure}[t]
	\begin{center}
		\centerline{\includegraphics[width=1.1\columnwidth]{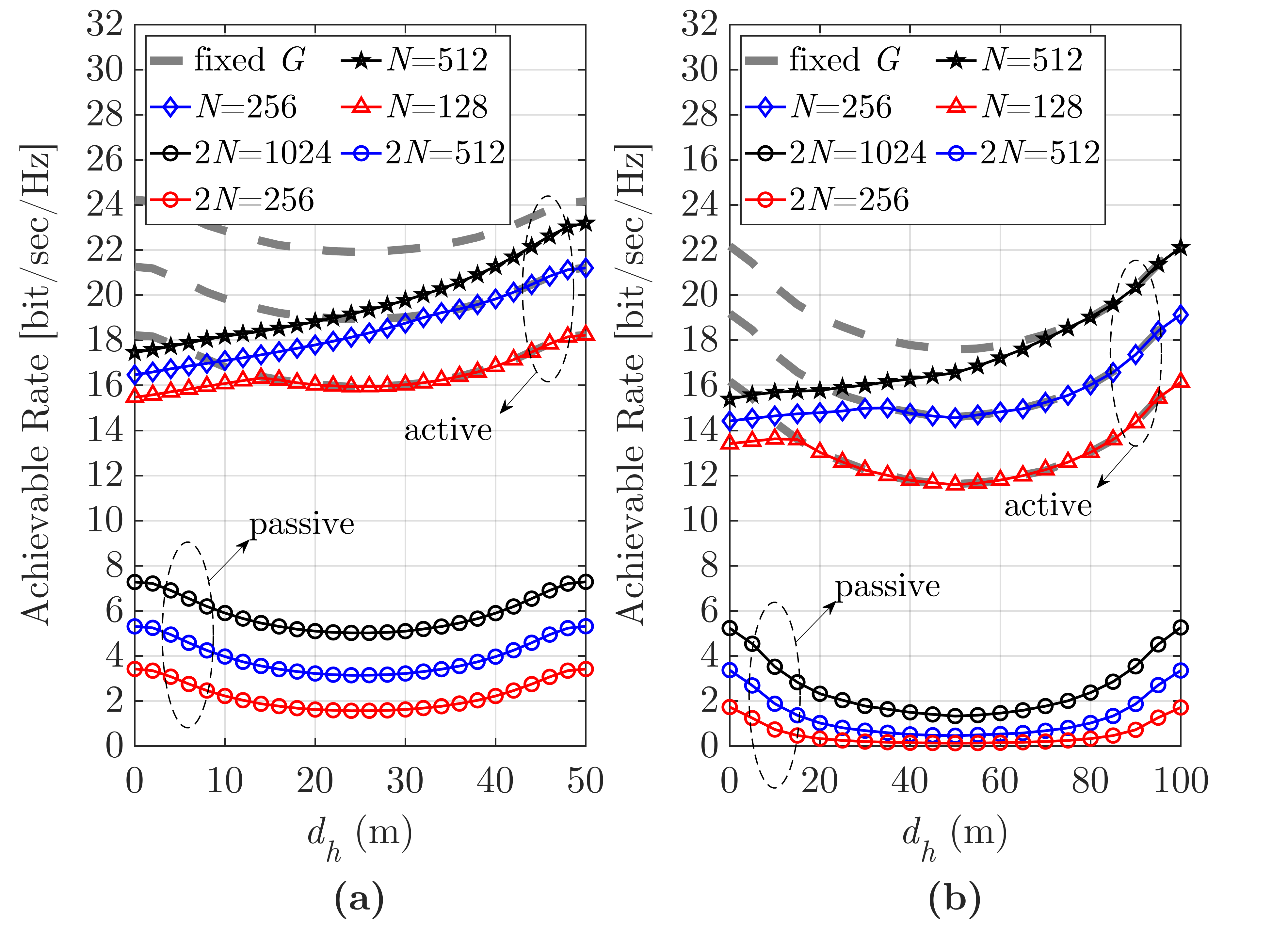}}
		\caption{\textbf{Achievable rates of the system for (a) \boldmath$d_h=50$ m and (b) \boldmath$d_h=100$ m.}}
		\label{fig:Fig4}
	\end{center} \vspace*{-0.7cm}
\end{figure}

Achievable rates of both designs for varying $d_h$ with several $N$ values are presented in Fig. \ref{fig:Fig4}. For this figure, the lines labeled as $N=128$, $256$, and $512$ represent the achievable rates for the active model, where those labeled with $2N=$ $256$, $512$, and $1024$ stand for the passive model. The dashed lines show the performance behaviours when there is not any output power limitation for the amplifier (i.e., results with a fixed $G$ value of $30$ dB). In Fig. \ref{fig:Fig4}(a), the achievable rates for the active model overlap with the corresponding dashed lines when $P_\text{out} \leq P_\text{max}$ and $G_\text{opt}=G_\text{max}$. We observe that the intersection points with the dashed lines occur at smaller values of $d_h$ as $N$ decreases. The reason behind is that $G_\text{opt}$ begins to decrease after the intersection points as the amplifying RIS gets closer to the Tx to keep $P_\text{out}=P_\text{max}$. Moreover, it is a well known fact that placing an RIS close to the Tx or Rx results in a better system performance. This can be seen in both the Figs. \ref{fig:Fig4}(a) and (b), however, that is not the case with the amplifying RIS design. Placing the amplifying RIS close to the Tx provides lower achievable rates. We mentioned earlier that the reason behind this is the limitation of $P_\text{out}$ and having a greater $P_\text{max}$ would be a solution to this problem. Based on these examinations, one can conclude that the amplifying RIS can overcome the common problem that generally affects system performance unpleasingly in passive RIS designs, which is the necessity of placing the RIS close to the Tx or Rx. As seen in Fig. \ref{fig:Fig4}(a), the worst achievable rate is obtained when the RIS is placed in the middle of the Rx and Tx for the passive RIS. This gets even worse in Fig. \ref{fig:Fig4}(b) when the Tx and Rx are further apart. The amplifying RIS design significantly reduces the multiplicative path loss effect by amplifying the combined signal by RIS\textsubscript{1} as it uses a PA between the two RISs. Using the amplifying RIS instead of the passive one can compensate this performance drop, even enhance the performance further.
\begin{figure}[t]
	\begin{center}
		\centerline{\includegraphics[width=1\columnwidth]{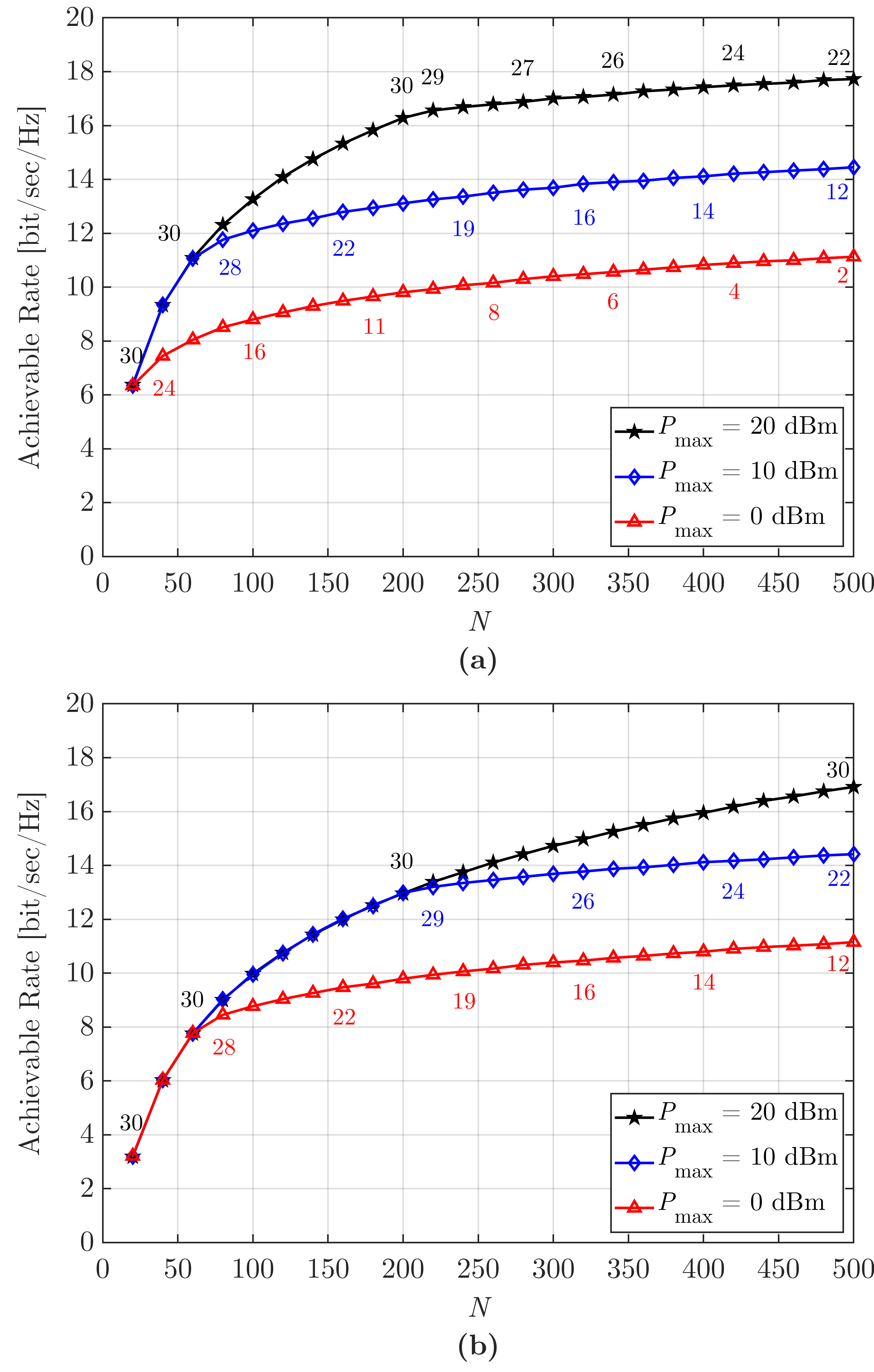}}
		\caption{\textbf{Achievable rates of the amplifying RIS-assisted system for (a) \boldmath$P_t=20$ dBm and (b) \boldmath$P_t=10$ dBm. The numbers on the markers indicate \boldmath$G_\text{opt}$ values.}}
		\label{fig:Fig5}
	\end{center} \vspace*{-0.6cm}
\end{figure}

Similar to passive RIS scenarios, the RIS size has a considerable impact on achievable rate also for the active RISs. Fig. \ref{fig:Fig5} illustrates the achievable rates of the amplifying RIS design where Figs. \ref{fig:Fig5}(a) and (b) stand for $P_t=20$ dBm and $P_t=10$ dBm, respectively. Here, the numbers for each marker signify $G_\text{opt}$ for corresponding $N$. Here, $G_\text{opt}$ begins to decrease after a point for all of the cases except for $P_\text{max}=20$ dBm in Fig. \ref{fig:Fig5}(b). Different from the previous case, the reason behind the occurrence of the break points is the bigger $N$ values, such that $P_\text{in}$ increases if more reflecting elements are used and again limits $G_\text{opt}$. In Fig. \ref{fig:Fig5}(a), $P_\text{out}$ reaches $P_\text{max}$ for large $N$ and amplifier cannot operate with $G_\text{max}$ even if $P_\text{max}=20$ dBm. However, when $P_t$ is reduced to 10 dBm as in Fig. \ref{fig:Fig5}(b), we see that $P_\text{out}$ cannot reach $P_\text{max}=20$ dBm even for large $N$ and the amplifier can boost the signal with $G_\text{max}$. 
\begin{figure}[t]
	\begin{center}
		\centerline{\includegraphics[width=1.05\columnwidth]{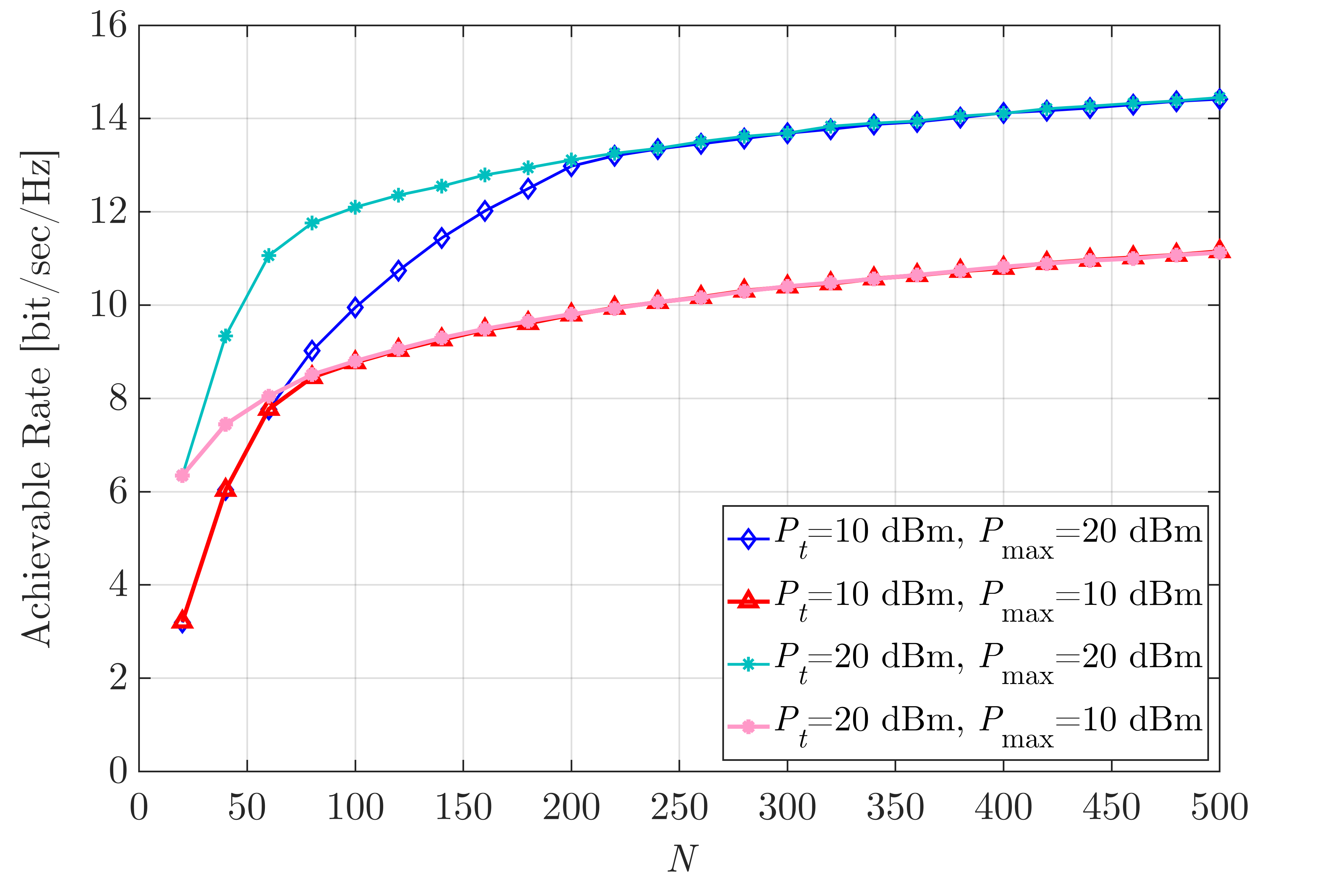}}
		\caption{\textbf{Achievable rates of the amplifying RIS-assisted system for different \boldmath$P_t$ and \boldmath$P_\text{max}$.}}
		\label{fig:Fig6}
	\end{center} \vspace*{-0.6cm}
\end{figure}

Tx power and RIS size cause similar effects on the system performance because both of them enhance $P_\text{in}$. As seen from Figs. \ref{fig:Fig5}(a) and (b), increasing $P_t$ does not always have an extra constructive effect on the system capacity because the achievable rates become identical for the same $P_\text{max}$ after a certain point as clearly demonstrated in Fig. \ref{fig:Fig6} as well. In Fig. \ref{fig:Fig6}, the achievable rates are the same after $N=200$ and $N=60$ when $P_\text{max}=20$ dBm and $P_\text{max}=10$ dBm, respectively, because $P_\text{out}=P_\text{max}$ at these $N$ values. Since $P_\text{max}=10$ dBm can be reached by lower $P_\text{in}$ values, the intersection point occurs at lower $N$ compared to the case $P_\text{max}=20$ dBm. On the other hand, a considerable performance difference is examined where $N$ is smaller than 100 and 50 for $P_\text{max}=20$ dBm and $P_\text{max}=10$ dBm, respectively. These observations indicate that the system performance is not enhanced with $P_t$ after the break points, nevertheless, it can be boosted slightly by using larger $N$. This can be explained by the fact that there are two RISs at both sides of the amplifier. Increasing $N$ for RIS\textsubscript{1} can enhance $P_\text{in}$ just as $P_t$, however, $P_\text{out}$ will be the same if it is already at $P_\text{max}$, so this increment in $N$ for RIS\textsubscript{1} does not affect the achievable rate. Despite this, if more reflecting elements are used for RIS\textsubscript{2}, slightly higher achievable rates are achieved although the total power is the same such that it is distributed among those reflecting elements. The reason of this advance is the beamforming gain which increases with $N$.

Considering the results presented in this subsection, the main performance limiting factors are the PA parameters, which are $P_\text{max}$ and $G_\text{max}$. Therefore, $N$ and $P_t$ has a considerable effect on the performance depending on the PA parameters.
\begin{figure}[t]
	\begin{center}
		\centerline{\includegraphics[width=1.05\columnwidth]{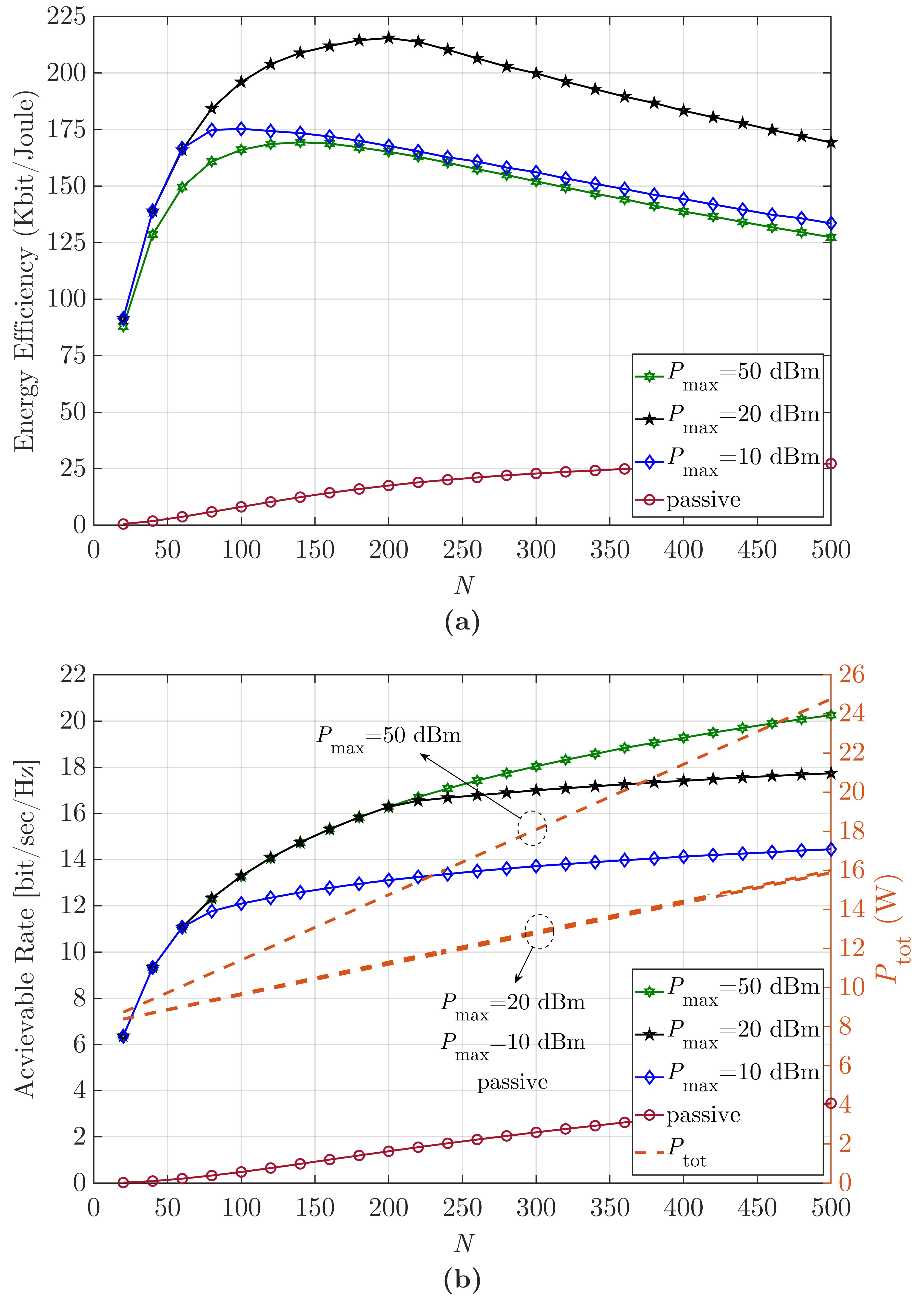}}
		\caption{\textbf{(a) Energy efficiencies of the amplifying RIS-assisted system for varying \boldmath$N$ and (b) corresponding achievable rates and \boldmath$P_\text{tot}$ values}}
		\label{fig:Fig7}
	\end{center} \vspace*{-0.7cm}
\end{figure}

\subsection{Energy Efficiency Evaluation}

In this subsection, numerical results for EE and power consumption are presented. Fig. \ref{fig:Fig7}(a) shows the EE values for varying $N$ and different $P_\text{max}$ values where Fig. \ref{fig:Fig7}(b) stands for the corresponding achievable rates and $P_\text{tot}$. In Fig. \ref{fig:Fig7}(a), the EE values start to decline after some points because the achievable rates are almost at a constant level for $P_\text{max}=20$ dBm and $P_\text{max}=10$ dBm while $P_\text{tot}$ is still increasing as seen in the Fig. \ref{fig:Fig7}(b). When $P_\text{max}=50$ dBm, a worse EE performance is obtained as the increase in power consumption has a greater effect than the increase in the achievable rate. Unless $P_\text{max}$ is too high, the total power consumption of the active design is nearly the same as the passive RIS. Consequently, the active design performs better in terms of EE.

\begin{figure}[t]
	\begin{center}
		\includegraphics[width=1\columnwidth]{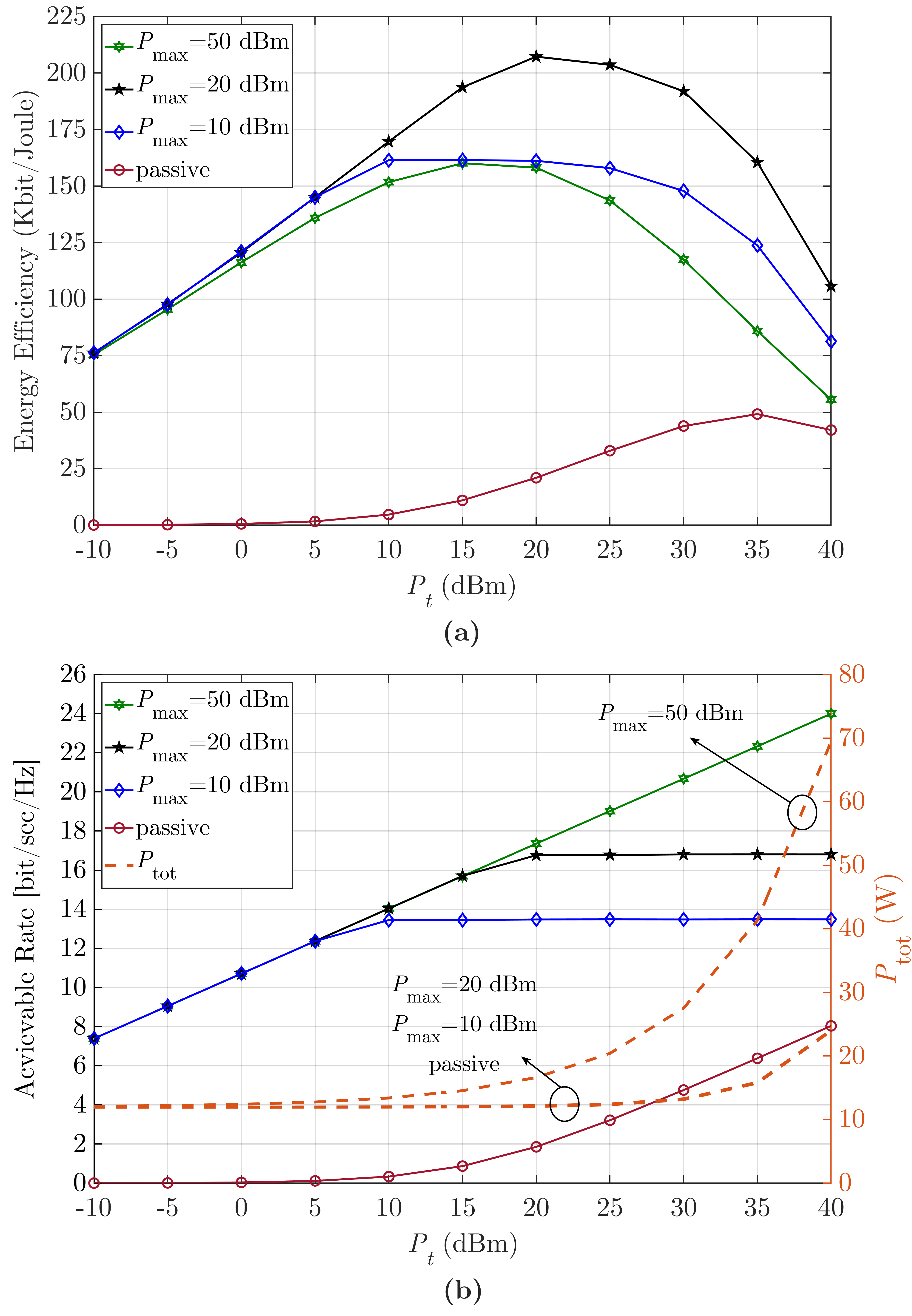}
		\caption{\textbf{(a) Energy efficiencies for different \boldmath$P_\text{max}$ values varying with \boldmath$P_t$ and (b) corresponding achievable rates and \boldmath$P_\text{tot}$ values}}
		\label{fig:Fig8}
	\end{center} \vspace{-0.3cm}
\end{figure}

EE is investigated for varying $P_t$ and different $P_\text{max}$ values in Fig. \ref{fig:Fig8}(a). Achievable rates and $P_\text{tot}$ values are also given in  Fig. \ref{fig:Fig8}(b) for the same scenario. For $P_\text{max}=20$ dBm and $P_\text{max}=10$ dBm, although the amplifying RIS does not consume high amount of power, the EE begins to decrease after some points. The reason is that increasing $P_t$ after these points does not affect the achievable rate, but increases $P_\text{tot}$. For the case of $P_\text{max}=50$ dBm, as we do not reach $P_\text{max}$, both the power consumption of the PA at the base station and the PA between the RISs increases. Accordingly, we observe that the EE starts to decrease after a certain point even if the achievable rate keeps increasing. 

\begin{figure}[t]
	\begin{center}
		\includegraphics[width=1\columnwidth]{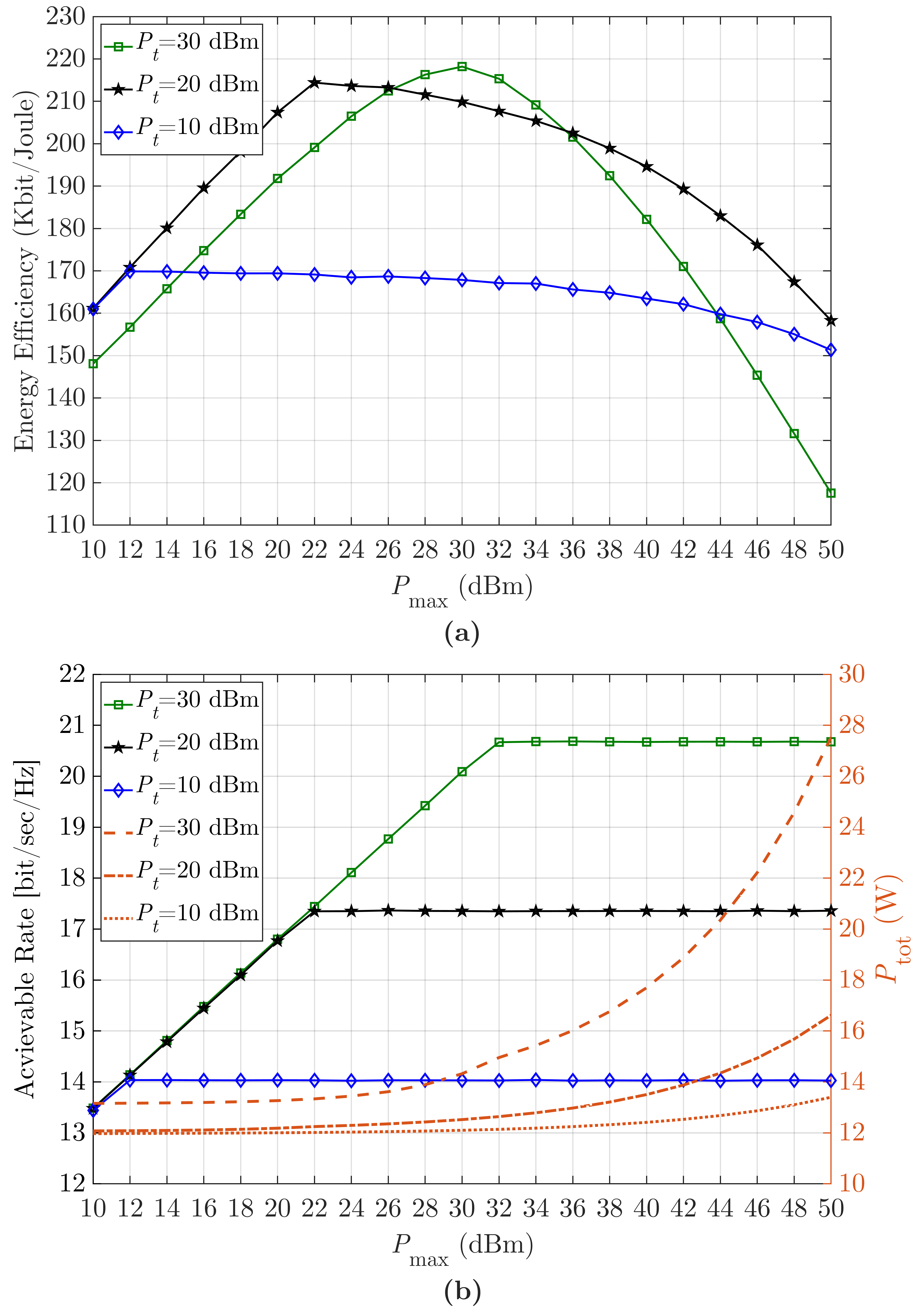}
		\caption{\textbf{(a) Energy efficiencies for different \boldmath$P_t$ values varying with \boldmath$P_\text{max}$ and (b) corresponding achievable rates and \boldmath$P_\text{tot}$ values}}
		\label{fig:Fig9}
	\end{center} \vspace*{-0.31cm}
\end{figure}

Finally, Figs. \ref{fig:Fig9}(a) and (b) represent the EE, achievable rate, and $P_\text{tot}$ for varying $P_t$ and different $P_\text{max}$ values. Unlike Figs. \ref{fig:Fig7} and \ref{fig:Fig8}, the break points occur when the amplifier cannot amplify the signal more than $G_\text{max}=30$ dB. Thus, increasing the $P_\text{max}$ only leads to higher power consumption as explained in (\ref{eq:pamp}) while it does not have a positive effect on the achievable rate. This causes the EE to show an interesting downward trend.

In the view of these observations, one can conclude that the EE is affected by many parameters such as $P_t$, $P_\text{max}$, $N$, and $G_\text{max}$. As shown in \eqref{eq:pamp}, $P_\text{max}$ and $P_t$ have a more clear effect on the EE compared to the other parameters. Increasing these parameters to have a higher achievable rate does not always lead to a more energy efficient system. To reach a system with a maximum EE, all of these parameters should be jointly optimized, which requires advanced optimization techniques and might be a topic for the future works.

\section{Conclusion}
In this paper, we have proposed an amplifying RIS design that utilizes a single PA. The differences and advantages of the proposed design have been clearly pointed out compared to the conventional passive RIS design. We have presented its signal model for the amplifying RIS systems, and optimized the amplifier gain and the phase response of RIS\textsubscript{1} and RIS\textsubscript{2} to maximize the achievable rate. Ultimately, active RIS-assisted systems provide higher communication capacity and better error rate performance than the passive RIS-assisted systems. In addition, they enable the flexibility to place the amplifying RIS anywhere in between Tx and Rx as it greatly reduces the effect of double path loss, unlike passive RIS that should be placed closer to Tx or Rx. On top of that, active RIS-assisted systems are more energy efficient, although they consume more power. 

This paper presented a new design for active RIS architectures by deploying a single active component. Extending our design to MIMO systems introduces complex optimization problems such as the joint optimization of transmit precoding, phases of the reflecting elements, and gain of the PA needed. Potentially, deep learning and alternating optimization algorithms can be used to deal with these problems, however, these are beyond the scope of this first study. Future works may also include joint uplink and downlink communication, the optimization and extension of this system to MIMO systems, as well as real-world experimental results.

\bibliographystyle{IEEEtran}
\bibliography{IEEEabrv,ref}


%

\begin{IEEEbiography}[{\includegraphics[width=1in,height=1.25in,clip,keepaspectratio]{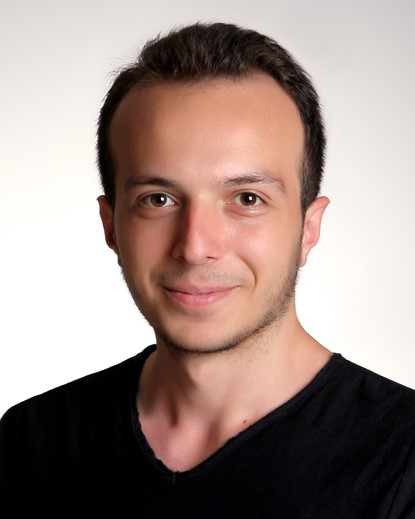}}]{\textbf{Recep A. Tasci}} (S'20) received his B.S degree in Electrical and Electronics Engineering from Istanbul Medipol University in 2020. He is currently a research and teaching assistant at Koç University. His research interests include wireless communications, reconfigurable intelligent surfaces, channel modeling, and signal processing.
\end{IEEEbiography}
\begin{IEEEbiography}[{\includegraphics[width=1in,height=1.25in,clip,keepaspectratio]{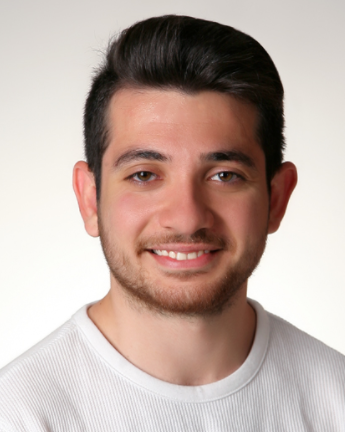}}]{\textbf{Fatih Kilinc}} (S'20) received his B.S degree from Istanbul Medipol University in 2020. He is currently pursuing M.S. degree at Koç University. He is a research and teaching assistant at Koç University. His research interest include channel modeling, intelligent surfaces and signal processing for wireless communications.
\end{IEEEbiography}
\begin{IEEEbiography}[{\includegraphics[width=1in,height=1.25in,clip,keepaspectratio]{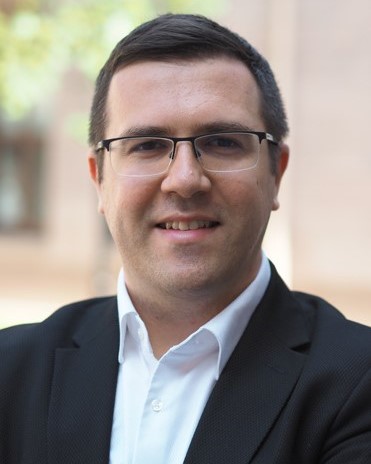}}]{\textbf{Ertugrul Basar}} (S'09-M'13-SM'16) received his Ph.D. degree from Istanbul Technical University in 2013. He is currently an Associate Professor with the Department of Electrical and Electronics Engineering, Koç University, Istanbul, Turkey and the director of Communications Research and Innovation Laboratory (CoreLab). His primary research interests include beyond 5G systems, index modulation, intelligent surfaces, waveform design, and signal processing for communications. Dr. Basar currently serves as a Senior Editor of IEEE Communications Letters and an Editor of IEEE Transactions on Communications and Frontiers in Communications and Networks. He is a Young Member of Turkish Academy of Sciences and a Senior Member of IEEE.
\end{IEEEbiography}
\begin{IEEEbiography}[{\includegraphics[width=1in,height=1.25in,clip,keepaspectratio]{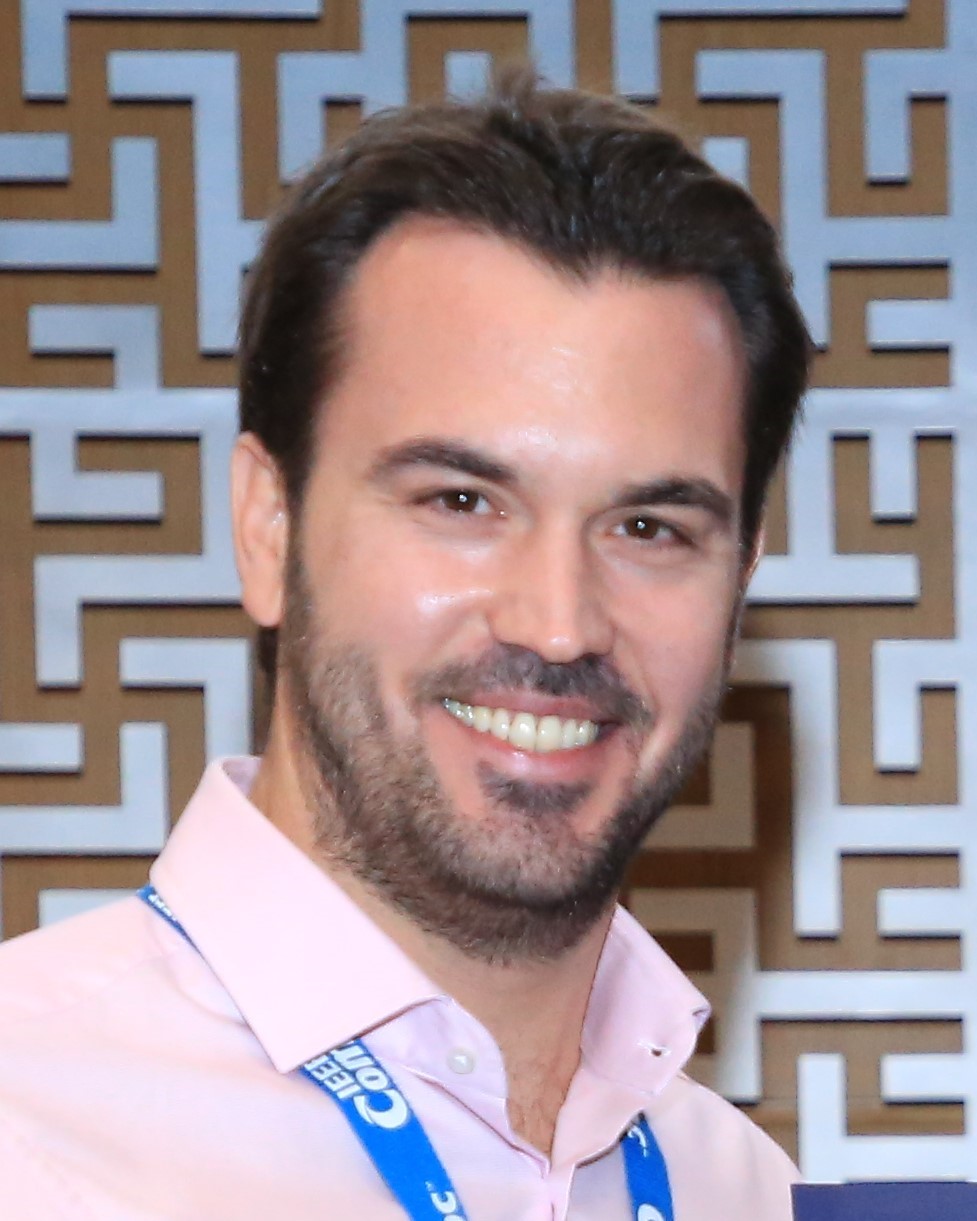}}]{\textbf{George C. Alexandropoulos}} (S'07-M'10-SM'15) is an Assistant Professor with the Department of Informatics and Telecommunications, National and Kapodistrian University of Athens, Greece. His research interests span the general areas of algorithmic design and performance analysis for wireless networks with emphasis on multi-antenna transceiver hardware architectures, active and passive reconfigurable metasurfaces, integrated communications and sensing, millimeter wave and THz communications, as well as distributed machine learning algorithms. He has received the best Ph.D. thesis award 2010, the IEEE Communications Society Best Young Professional in Industry Award 2018, the EURASIP Best Paper Award of the Journal on Wireless Communications and Networking 2021, the IEEE Marconi Prize Paper Award in Wireless Communications 2021, and a Best Paper Award from the IEEE GLOBECOM 2021.
\end{IEEEbiography}


\end{document}